\documentstyle[prd,aps,twocolumn,epsf]{revtex}
\def\br{\begin{eqnarray}}
\def\er{\end{eqnarray}}
\def\be{\begin{equation}}
\def\ee{\end{equation}}

\def\({\left(}
\def\){\right)}

\begin{document}
\twocolumn[\hsize\textwidth\columnwidth\hsize\csname 
@twocolumnfalse\endcsname                            
%
%
\title{Can we observe the $\sigma$ meson in peripheral heavy ion collisions?}
\author{
A.~A.~Natale and  C.~G.~Rold\~ao\\
}
\address{
Instituto de F\'{\i}sica Te\'orica,
Universidade Estadual Paulista,
Rua Pamplona 145,
01405-900, S\~ao Paulo, SP,
Brazil}
\date{\today}
\maketitle
\begin{abstract}
We discuss the possible production of the $\sigma$ meson and detection of
its two photon decay mode in peripheral heavy ion collisions at RHIC. Assuming
the mass and total decay width values reported recently by the E791 Collaboration,
we verify that the $\sigma$ meson can be observed in the subprocess $\gamma \gamma
\rightarrow \gamma \gamma$ if its partial decay width into photons is of
the order of a few keV.
\end{abstract}

\pacs{PACS: 25.75.-q, 27.75.Dw, 13.60.-r, 14.40.C}

\vskip 0.5cm]

The possible existence of a light scalar sigma ($\sigma$) meson
has been a controversial subject for roughly forty
years\cite{sigma}. In the strong interaction phenomena the
$\sigma$ meson plays a role analogous to the Higgs particle in the
Weinberg-Salam theory. It is expected that when the chiral
symmetry of the strong force between quarks is broken by the
nonperturbative vacuum, generating a massless pseudoscalar meson
(the pion), a massive scalar meson (the $\sigma$) is also
generated\cite{nambu}. The sigma meson is fundamental for
effective theories describing the nuclear force, where it is
responsible for a attractive part of the nuclear
potential\cite{kun}. Although its experimental confirmation would
be quite important, its existence has remained inconclusive for a
long time making it disappear from the review of particle physics
and reappearing only recently\cite{sigma,RPP}. There are two
difficulties for its discovery: the extraction of the scalar
properties from experiment and the lack of knowledge about its
underlying quark substructure. Because the $J=0$ channel may
contain strong competing contributions, such resonance may not
necessarily dominate its amplitudes and could be hard to
``observe". In such an instance its verification would be linked
to the model used to its description.

One of the most compelling evidences for
the $\sigma$ meson was reported recently by the E791 Collaboration,
resulting from the observation of the $D^+ \rightarrow \pi ^- \pi ^+ \pi ^+$
decay \cite{e791}. The Dalitz-plot of this decay can hardly be
fitted without a $0^{++}$ ($\sigma$) resonance, whose
mass and total decay width are $478 ^{+24}_{-23} \pm 17$
MeV and $324^{+42}_{-40} \pm 21$ MeV, respectively. The $\sigma$
meson appears as an intermediate state decaying into two pions, which
is the same final state questioned over many years as a possible
(or not) signal for this broad resonance.

The $\sigma$ meson can also decay into two photons, although this
decay is orders of magnitude smaller than the $\pi \pi$ decay mode
it has the advantage that the final state is not strongly interacting.
Therefore, the observation of the decay $\sigma \rightarrow \gamma \gamma$
would be a cleaner way to establish its existence. This decay is also
important to unravel any possible amount of mixing of quarks or gluons
in this resonance\cite{close}. The ideal observation of this resonance would
be through the process $\gamma \gamma \rightarrow \sigma \rightarrow \gamma \gamma$,
where the production and decay modes are free of any problem caused by
the strong interaction of the initial or final states which may hide
its properties. Our work will focus exactly on the possibility of
observing this meson in peripheral heavy ion collisions through
the subprocess $\gamma \gamma \rightarrow \gamma \gamma$.

In the Relativistic Heavy-Ion Collider (RHIC), operating at the
Brookhaven Laboratory, there is a program for very peripheral heavy
ion collisions, where each charge $Z$ accelerated ion has a strong
photon field around it that may originate electromagnetic interactions,
and where the ions remain intact after these interactions. The huge
luminosity of photons in these collisions open the possibilities of
studying two-photon physics as discussed by several authors
\cite{baur}, and, in particular, the process $\gamma \gamma
\rightarrow \gamma \gamma$ could indeed be measured\cite{bertulani}.

We compute the $\gamma \gamma \rightarrow \gamma\gamma$ subprocess, for
the Au ion at RHIC energies, $\sqrt{s} = 200$ GeV/nucleon. The photon
distribution in the nucleus can be described using the
equivalent-photon or Weizs\"{a}cker-Williams approximation in the
impact parameter space. We use the photon distribution obtained
by Cahn and Jackson~\cite{cahn}, wich takes into account the condition
for realistic peripheral collisions (for details see Ref.\cite{we}).
The subprocess $\gamma \gamma \rightarrow \gamma \gamma $
up to energies of a few GeV is dominated by the continuous
fermion box diagram, and is a background for the resonant
$\gamma \gamma \rightarrow \sigma \rightarrow \gamma \gamma$ process.
It was first calculated exactly by Karplus
and Neuman \cite{karplus} and De Tollis \cite{tollis2}.
There are sixteen helicity amplitudes for the process and,
due to symmetry properties, the number of independent
amplitudes will be only five, that may be
chosen to be $M_{++++}$, $M_{++--}$, $M_{+-+-}$, $M_{+--+}$ and
$M_{+++-}$. Where the $+$ or $-$ denotes the circular
polarization values $+1$ and $-1$. The remaining helicity
amplitudes may be obtained from parity and permutation symmetry.
Of these five helicity amplitudes, three are related by crossing,
hence it is sufficient to give just three, which are presented in
detail in Ref.\cite{we}.

The electron gives the major contribution to the continuous
$\gamma \gamma\rightarrow \gamma\gamma$ subprocess. The muon and $u$
contributions are smaller roughly by one order of magnitude, and the
$d$ and $s$ quark contributions are even smaller. The process is
proportional to $(q_f ^2)^4$  where $q_f$ is their charge, and this
is the main reason for the suppression of the quarks contribution\cite{we}.
Contribution from pion loops, heavy quarks or leptons,
and charged weak bosons are negligible up to O(2 GeV).
The main background for this reaction could come from the
processes: a) Double-Pomeron exchange, this process was computed and after
extracting the contribution of central collisions (see Ref.\cite{eu} for
details of the procedure) they are shown to be negligible, in agreement
with the conclusions of our work about the importance of double Pomeron
contributions to ultra peripheral heavy ion collisions\cite{eu},
b) Double bremsstrahlung (which dominates the
region of $| \cos{\theta} | \approx 1$, where $\theta $ is the scattering
angle in the two photon center-of-mass system). This background is eliminated
with cuts in the scattering angle, and in all calculations we will consider
the maximum value $|\cos \theta | = 0.5$. This cut is also consistent
with all the requirements proposed by Nystrand and Klein\cite{klein}
to eliminate background, and retains only the back-to-back photons in
the central region of rapidity. However, a full simulation of
bremsstrahlung contributions should be performed to estimate
the effectiveness of this cut.
Note that the angular cut diminishes the contribution of the continuous
process (peaked at large angles) in comparison with the
isotropic photon distribution produced by a scalar resonance.
As discussed in Ref.\cite{bertulani}, the $\gamma \gamma$
scattering can indeed be measured in peripheral heavy ion
collisions. Just to give one
idea of the number of events, with a luminosity of 2.0 $\times
10^{26}$ cm$^{-2}$s$^{-1}$ \cite{klein} and integrating
in a bin of energy $700 \pm 100$ MeV (which is
free of any strong resonance decaying into two-photons), we have
1532 events/year assuming $100\%$ efficiency in the tagging of
the ions and photon detection\cite{we}.

Another possible background is the
continuous or resonant ($\sigma$) production of a pair of neutral
pions where both are misidentified with photons. This accidental
background can be easily isolated measuring its invariant mass
distribution, and making a cut that discriminates a single photon
from one pion that subsequently decays into two photons. For example,
in the case of the sigma meson each one of the neutral pions from its much
large hadronic decay should be misidentified. These pions would produce
pairs of photons with a large opening angle $\phi$, where
$\cos{(\phi /2)} = \sqrt{1 - 4m_{\pi}^{2}/m_{\sigma}^2}$.
However, the calorimeters already in use in many
experiments are able to distinguish between these two and single
photon events with high efficiency (see, for instance,
Ref.\cite{alde}). The detectors at RHIC are prepared
to detect photons of O($1$) GeV, because these are also a
signal for the quark-gluon plasma formed in central collisions.

The contribution from $\gamma-$Pomeron$\rightarrow {\rm V} \rightarrow \gamma \gamma
X$ may also produce another accidental background.  The vector mesons, V, are produced with a $p_T$
distribution similar to the $\sigma$ and at higher rates than
$\gamma \gamma \rightarrow {\rm scalar/tensor}$, {\it e.g.}\,
$\gamma-$Pomeron$\rightarrow \omega$ rate is 10 Hz at RHIC\cite{kn}, three
orders of magnitude higher than for a similar mass meson of spin
0 or 2. The $\omega$ branching ratio to three photons is $8.5\%$.
If a small $p_T$ photon from this decay is undetected, one is
left with a low $p_T$ two-photon final state with $M<780$ MeV
that could be taken for a $\sigma$ candidate.  This background
is likely to be much greater than the proposed signal.
Furthermore, the $\sigma$ is very wide, making background
subtraction difficult.  At higher masses, one also has $\phi
\rightarrow \eta \gamma$, $\pi^0 \gamma$, $K_L K_S \rightarrow
\gamma X$ as well as $\gamma \gamma \rightarrow f_2(1270)
\rightarrow \pi^0 \pi^0$ and possibly copious production of
$\rho(1450)$ and $\rho(1700)$ by $\gamma-$Pomeron interactions.
Clearly a full simulation of all these background processes
should be kept in mind when measuring two-photon final states.

\begin{figure}[htb]
\epsfxsize=.35\textwidth
\begin{center}
\leavevmode \epsfbox{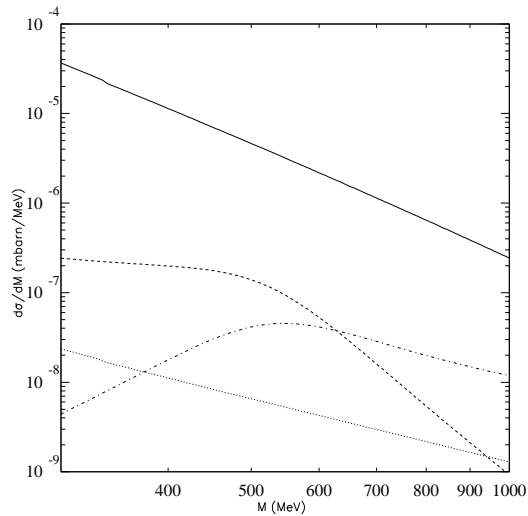}
\end{center}
\vskip -1.0cm
 \caption{ Invariant mass distribution of photon pair production.
 The solid curve is due to box diagram only, the dashed one is due to the process
 $\gamma \gamma \rightarrow \sigma \rightarrow \gamma \gamma$ in the
 Breit-Wigner approximation, the dash-dotted is the scalar contribution
 of Eq.(\ref{escalar}), and the dotted one is the continuum Pomeron
 process $P P \rightarrow \gamma \gamma$. We assumed $\Gamma_{\gamma\gamma} = 4.7$ keV,
 and the angular cut is equal to $-0.5 < \cos \theta < 0.5$ .  }
  \label{interferencia_escalar}
\end{figure}

Photon pair production via the box diagram is a background to
$\gamma \gamma \rightarrow \sigma \rightarrow \gamma \gamma $ process
(or vice versa), both have the same initial and final states, and
for this reason they can interfere one in another. Normally the
interference between a resonance (R) and a continuum process is
unimportant, because on resonance the two are out of phase.
Assuming a Breit-Wigner profile the total cross section for the
elementary subprocess $\gamma
\gamma \rightarrow R \rightarrow \gamma \gamma $ is
\begin{eqnarray}
\frac{d \sigma ^{\gamma \gamma}_{ZZ}}{dM} = 16 \pi \frac{dL}{dM}
\frac{\Gamma ^2 (R \rightarrow \gamma \gamma)} {(M^2-m_R^2)^2
+m_R^2 \Gamma^2_{total}},
 \label{dsigfoton}
\end{eqnarray}

 \noindent
 where $L$ is the photon luminosity, $M$ is the energy of the photons
 created by the collision of the ions. $\Gamma (R\rightarrow \gamma \gamma
)$($\equiv \Gamma_{\gamma \gamma}$) and $\Gamma_{total}$ are the
partial and total decay width of the
 resonance with mass $m_R$ in its rest frame. The results for the
 box diagram and the $\sigma$ production in the Breit-Wigner approximation
 are shown in Fig.(\ref{interferencia_escalar}). We used the value
$\Gamma_{\gamma\gamma} = 4.7$ keV determined by Pennington and
Boglione\cite{pennington}. This value is in the middle of various
determinations (for other values see Ref.\cite{we}). The
$\sigma$ mass and total decay width correspond to the central values of
Ref.\cite{e791}. The use of a constant total width in the $\sigma$ resonance
shape is not always a good approximation\cite{schechter}. We
discuss the $\gamma \gamma \rightarrow \gamma \gamma$ process
above the two pions threshold where the peculiarities of the broad
resonance, basically due to the $\sigma$ decay into pions, are not
so important. However, we also computed the
cross section with a energy dependent total width
$\Gamma(M) \simeq \Gamma_0 \left( {p^*}/{p^*_0} \right) ^{2J + 1}$,
where $p^* = p^*(M)$ is the momentum of decay particles at mass $M$,
measured in the resonance rest frame, $p^*_0 = p^*(m_R)$,
which, as shown by Jackson many years ago\cite{jackson}, is more
appropriate for a quite broad resonance. The net effect is a
slight distortion of the cross section shape with a small increase
of the total cross section, and we shall not consider it again.

Off resonance we can expect a negligible contribution for the
process $\gamma \gamma \rightarrow R \rightarrow \gamma \gamma $
and consequently the same for its interference with the continuum
process. This is true if the resonance has a small total decay
width\cite{we}, but this is not the $\sigma$ case. To take into account
the interference we must make use of a model to calculate the helicity
amplitudes of the $\sigma$ meson exchange. Using the effective
lagrangian $g_s F^{\mu \nu} F_{\mu \nu} \Phi_s $,
where $g_s$ is the coupling of the photons to the scalar field
$\Phi_s$ and $F^{\mu \nu}$ is the electromagnetic field tensor
the following amplitudes comes out\cite{we,tollis3}:
\begin{eqnarray}
M_{++++} &=& - \frac{2 \pi}{\alpha^2} F(\lambda r_t),
 \nonumber \\
M_{+-+-} &=& - \frac{2 \pi}{\alpha^2} F(\lambda t_t),
 \nonumber \\
M_{+--+} &=& - \frac{2 \pi}{\alpha^2} F(\lambda s_t),
 \nonumber \\
 M_{+++-} &=& 0,
  \nonumber \\
 M_{++--} &=& - \frac{2 \pi}{\alpha ^2} \{ F(\lambda s_t) + F(\lambda
 t_t) + F(\lambda r_t) \} .
\label{escalar}
\end{eqnarray}
where $\alpha $ is the fine-structure constant, $\lambda
= (m_f/m_R)^2$, and
\begin{eqnarray}
F(x) = 16 x^2\frac{\Gamma_{\gamma \gamma}}{m_R} \left( 4 x - 1 + i
\frac{\Gamma _{total}}{m_R} \right)^{-1},
 \label{fun_res}
\end{eqnarray}
and  $r_t$, $s_t$ and $t_t$ are related with the standard
 Mandelstam variables s, t, and u by
$r_t = \frac{1}{4} \frac{s}{m_f^2}$, $s_t = \frac{1}{4}
\frac{t}{m_f^2}$ and $t_t = \frac{1}{4} \frac{u}{m_f^2}$.

These amplitudes and the ones describing the fermion (with mass
$m_f$) box diagram enter in the expression for the differential cross
section of photon pair production from photon fusion
${d \sigma}/{d \cos \theta }= ({1}/{2 \pi}) ({\alpha
^4}/{s}) (\sum_f q^2_f)^4 \sum |M|^2$,
to give the total cross section of Fig.(\ref{interferencia_escalar}).
We verified that the interference is destructive. The continuum
process ($P P \rightarrow \gamma \gamma$) is also shown in
Fig.(\ref{interferencia_escalar}). The diffractive resonant event is
even smaller and is not shown in this figure.
\vskip -1.0cm
\noindent

The effective Lagrangian model used to compute the $\sigma$
contribution to the photon pair production gives a larger
cross section than the calculation with the Breit-Wigner approximation
at energies above $M \approx 600$ MeV. It is dominated
by the $s$ channel contribution. We consider the Breit-Wigner result as the
best signal representation for the resonant process because we are
using the E791 data and this one was fitted by a Breit-Wigner profile.
The effective Lagrangian gives a nonunitary amplitude that
overestimates the sigma production above $600$ MeV and shows
the model dependence in the $\sigma$ analysis that we commented before.
The Breit-Wigner profile is not a bad
approximation as long as we stay above the two pions threshold and in the following
we assume that the signal is giving by it
(the dashed curve of Fig.(\ref{interferencia_escalar})) and the background is giving by
the box diagram result (the solid curve of Fig.(\ref{interferencia_escalar})). Note that, due to the
destructive interference, the actual measurement will give a curve
below the solid curve of Fig.(\ref{interferencia_escalar}).

From the experimental point of view we would say that the reaction
$\gamma \gamma \rightarrow \gamma \gamma$ has to be observed and
any deviation from the continuum process must be carefully
modeled until a final understanding comes out, with the
advantage that the final state is not strongly interacting.
Note that in this modeling the $\eta$ meson will contribute to
$\gamma \gamma \rightarrow \gamma \gamma$ in a small region of
momentum\cite{we}, even so it has to be subtracted in order to
extract the complete $\sigma$ signal.

\vskip -1.0cm
\begin{figure}[htb]
\epsfxsize=.35\textwidth
\begin{center}
\leavevmode \epsfbox{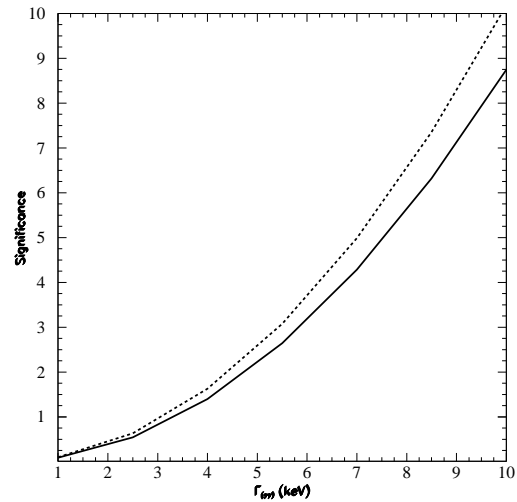}
\end{center}
\vskip -1.0cm
\caption{
Significance as a function of decay width into two photons,
 $\Gamma _{\gamma \gamma}$, for a sigma meson with mass equal to 478 MeV and
 total decay width of 324 MeV. The solid curve was obtained integrating
 the cross sections in the interval $438<M<519$ MeV, the dashed one in
 the interval $300<M<800$ MeV. The signal and background are giving by the
dashed (Breit-Wigner) and solid (box diagram) curves of
Fig.(\ref{interferencia_escalar}) respectively. }
  \label{significancia}
\end{figure}
\noindent

We changed the values of the $\sigma$ mass and total width around
the central ones reported by the E791 Collaboration. We do not
observed large variations in our result, but noticed that it is
quite sensitive to variations of the partial decay width into
photons. It is interesting to look at the values of the
significance which is written as $ {\cal
L}\sigma_{signal}/\sqrt{{\cal L}\sigma_{back}} $ and
characterizes the statistical deviation of the number of the
observed events from the predicted background. The significance
as a function of
the two photons decay width of the sigma meson,
with mass equal to 478 MeV and total decay width of 324 MeV, is
shown in Fig.(\ref{significancia}), were we used a luminosity of
${\cal L} = 2.0 \times 10^{ 26}$ cm$^{-2}$s$^{-1}$ at RHIC and
assumed one year of operation. The significance is above
$2\sigma$ $95\%$ confidence level limit for two photon decay
width greater than 4.7 keV, while for a $5\sigma$ discovery
criteria can be obtained with $\Gamma_{\gamma \gamma} > 7.5$ keV.
The numbers in Fig.(\ref{significancia}) were computed with the
signal given by the Breit-Wigner profile, and the background by
the pure box diagram. The solid curve was obtained integrating
the cross sections in the range of experimental mass uncertainty
$438<M<519$ MeV, while the dashed curve resulted
from the integration in the interval $300<M<800$ MeV. Note that
there is no reason, a priori, to restrict the
measurement to a small bin of energy. This choice will depend
heavily on the experimental conditions. Therefore, for
values of $\Gamma_{\gamma\gamma}$ already quoted in the
literature the sigma meson has a chance to be seen in its two
photon decay mode. The discovery limits discussed above refer
only to a statistical evaluation. Our work shows the importance
of the complete simulation of the signal and background including
an analysis of possible systematic errors that may decrease the
significance.

In summary, we discussed the possibility of observing the subprocess
$\gamma \gamma \rightarrow \gamma \gamma$ in peripheral heavy ion
collisions at RHIC. This process is important per se, but it is more
interesting because it can be intermediated by a scalar ($\sigma$)
resonance. The main advantage of this process is that the initial
and final states are not strongly interacting, providing a much cleaner
signal than the usually studied two pion decay of the $\sigma$ meson.
We computed the resonant process, the continuum one and discussed other
possible background. The resonant was determined within the Breit-Wigner approximation
and with an effective Lagrangian describing the $\sigma$ interaction
with photons. We determined a cut that leaves the cross sections
apart by one order of magnitude when using a partial decay width
of the sigma ($\Gamma_{\gamma\gamma}$) into two photons of 4.7 keV. Larger
and smaller values for $\Gamma_{\gamma\gamma}$ can be found in the
literature. The significance for the $\sigma$ discovery in this
reaction was determined and the observation is possible with one
year of data acquisition. It is important to perform a complete simulation
of the signal and background including an analysis of possible systematic
errors that may occur at the RHIC detectors.

\section*{Acknowledgments}
We would like to thank A.~C.~Aguilar, I.~Bediaga, C.~Dib, M.~M.~Leite,
R.~Rosenfeld and A.~Zimerman for many valuable discussions. This research
was supported by the Conselho Nacional de Desenvolvimento Cient\'{\i}fico e
Tecnol\'ogico (CNPq) (AAN), by Fundac\~ao de Amparo \`a Pesquisa
do Estado de S\~ao Paulo (FAPESP) (CGR,AAN) and by Programa
de Apoio a N\'ucleos de Excel\^encia (PRONEX).

\end{document}